\documentclass[preprint,12pt]{elsarticle}

\usepackage{amssymb}
\usepackage{epsfig}
\usepackage{graphics}
\usepackage{subfigure}
\usepackage{graphicx}

\journal{Physics Letters B}

\begin{document}

\begin{frontmatter}



\title{Observation of a first $\nu_\tau$ candidate in the OPERA experiment in the CNGS beam}

\author[label31]{N.~Agafonova}
\author[label33]{A.~Aleksandrov}
\author[label38]{O.~Altinok}
\author[label19]{M.~Ambrosio }
\author[label34]{A.~Anokhina}
\author[label26]{S.~Aoki}
\author[label35]{A.~Ariga}
\author[label35]{T.~Ariga}
\author[label5]{D.~Autiero}
\author[label36]{A.~Badertscher}
% ...

\fntext[korea]{Now at Chonnam National University, Korea}
\fntext[orlova]{Deceased}
\fntext[fn1]{Previously at LAL, Universit\'e de Paris 11, CNRS/IN2P3, 
             F-91898 Orsay, France}
\fntext[dellariccia]{Partially funded by Fondazione della Riccia}             
\fntext[berna]{Now at University of Bern, Switzerland}

\cortext[ca]{Corresponding Author}
\address[label31]{INR-Institute for Nuclear Research of the Russian Academy of Sciences, RUS-327312 Moscow,  Russia}
\address[label33]{LPI-Lebedev Physical Institute of the Russian Academy of Sciences, RUS-119991 Moscow, Russia}
\address[label38]{METU-Middle East Technical University, TR-06532 Ankara, Turkey}
\address[label19]{INFN Sezione di Napoli, I-80125 Napoli, Italy}
\address[label34]{(MSU SINP) Lomonosov Moscow State University Skobeltsyn Institute of Nuclear Physics , RUS-329992 Moscow, Russia}
\address[label26]{Kobe University, J-657-8501 Kobe, Japan}
\address[label35]{Albert Einstein Center for Fundamental Physics, Laboratory for High Energy Physics (LHEP), University of Bern, CH-3012 Bern, Switzerland}
\address[label5]{IPNL, Universit\'e Claude Bernard Lyon I, CNRS/IN2P3, F-69622 Villeurbanne, France}
\address[label36]{ETH Zurich, Institute for Particle Physics, CH-8093 Zurich, Switzerland}
% ...

\begin{abstract}
The OPERA neutrino detector in the underground Gran Sasso Laboratory (LNGS) has been designed to perform the first detection of neutrino oscillations in direct appearance mode through the study of the $\nu_\mu\rightarrow\nu_\tau$ channel. The hybrid apparatus consists of an emulsion/lead target complemented by electronic detectors and it is placed in the high energy long-baseline CERN to LNGS beam (CNGS) 730 km away from the neutrino source. Runs with CNGS neutrinos were successfully carried out in 2008 and 2009. After a brief description of the beam, the experimental setup and the procedures  used for the analysis of the neutrino events, we describe the topology and kinematics  of a first candidate  $\nu_\tau$ charged-current  event  satisfying the kinematical selection criteria. The background calculations and their cross-check are explained in detail and the significance of the event is assessed.
\end{abstract}


\end{frontmatter}

% \linenumbers

\section{Introduction}
\label{intro}
Neutrino oscillations were anticipated nearly 50 years ago\cite{pontecorvo} but were observed much later \cite{SK}. Several experiments carried out in the last decades with atmospheric and accelerator neutrinos, as well as with solar and reactor neutrinos, have contributed to our present understanding of neutrino mixing (see e.g. \cite{review} for a review). 

Atmospheric neutrino oscillations, in particular, have been studied by the  Kamiokande \cite{K}, Super-Kamiokande \cite{SK}, MACRO \cite{macro} and SOUDAN2 \cite{soudan2} experiments. The CHOOZ \cite{chooz} and Palo Verde \cite{paloverde} reactor experiments excluded indirectly the $\nu_\mu\rightarrow\nu_e$ channel as the dominant process in the atmospheric sector. Two accelerator experiments have already confirmed the oscillation hypothesis in disappearance mode, K2K \cite{K2K}  and MINOS  \cite{MINOS}. However, the direct observation of flavour transition through the detection of the corresponding lepton has never been observed. Appearance of $\nu_\tau$  will prove unambiguously that $\nu_\mu\rightarrow\nu_\tau$ oscillation is the dominant transition channel at the atmospheric scale.  This is the main goal of the OPERA experiment \cite{operaold, proposal1, proposal2}  with the high energy long baseline CNGS \cite{CNGS} neutrino beam from CERN to the Gran Sasso Underground Laboratory of INFN. The observation of interactions of CNGS neutrinos in the OPERA detector have already been presented \cite{operafirst}.

In this paper, we report the observation of a first $\tau$-lepton candidate event, possibly produced by an oscillated $\nu_\tau$, consistent with the kinematical selection criteria  \cite{proposal1, proposal2}. We evaluate the expected numbers of signal and background events in the data sample analysed so far, the probability that the event is due to a background fluctuation and the statistical significance of the observation.

\section{The OPERA detector and the CNGS beam}
\label{chap2}
The challenge of the OPERA experiment is the detection of the short-lived $\tau$ lepton ($c\tau$ = 87 $\mu$m) produced in the charged-current (CC) interaction of $\nu_\tau$. This sets two requirements difficult to conciliate: a large target mass to collect enough statistics and a very high spatial accuracy to observe the $\tau$ lepton, the decay length of which being in the millimetre range at the CNGS beam energy.
In OPERA, neutrinos interact in a large mass target made of lead plates interspaced with nuclear emulsion films acting as high-accuracy tracking devices. This kind of detector is historically called Emulsion Cloud Chamber (ECC). It was successfully used to establish the first evidence for charm in cosmic rays interactions  \cite{niu} and in the DONUT experiment for the first direct observation of the $\nu_\tau$ \cite{donut}. 
OPERA is an hybrid detector \cite{operadetector}  made of a veto plane followed by two identical Super Modules (SM) each consisting of a target section of about 625 tons made of 75000 emulsion/lead ECC modules or ''bricks'', of a scintillator Target Tracker detector (TT) to trigger the read-out and localize neutrino interactions within the target, and of a muon spectrometer. A target brick consists of 56 lead plates of 1 mm thickness interleaved with 57 emulsion films and weighs 8.3 kg. Their thickness along the beam direction corresponds to about 10 radiation lengths. In order to reduce the emulsion scanning load, Changeable Sheets (CS) film interfaces have been used. They consist in tightly packed doublets of emulsion films glued to the downstream face of each brick. 

Charged particles from a neutrino interaction in a brick cross the CS and produce signals in the TT that allow the corresponding brick to be identified and extracted by an automated system.  Large ancillary facilities are used to bring bricks from the target up to the automatic scanning microscopes at LNGS and various laboratories in Europe and Japan \cite{ESS,SUTS}. Extensive information on the OPERA detector and ancillary facilities are given in \cite{operadetector,facilities}.

OPERA is exposed to the long-baseline CNGS $\nu_\mu$ beam \cite{CNGS}  from CERN to LNGS, 730 km away. The beam is optimized for the observation of $\nu_\tau$  CC interactions. The average neutrino energy is $\sim$17 GeV. The $\bar{\nu}_\mu$ contamination is 2.1\% in terms of interactions; the $\nu_e$ and  $\bar{\nu}_e$ contaminations are lower than 1\%, while the number of prompt $\nu_\tau$  is negligible.  With a total CNGS beam intensity of 22.5 10$^{19}$ protons on target (p.o.t.), about 2s300 neutrino events would be collected. The experiment should observe about 10 $\nu_\tau$  CC events for the present $\Delta m^2_{23}$ allowed region with a background of less than one event. In 2006 and 2007, first interactions were recorded during the commissioning phase of the CNGS. Significant amount of data were collected starting in 2008 [15]. We report here on results based on part of the data taken during 2008 and 2009 runs, which amounted respectively to 1.78 and $3.52\times10^{19}$ p.o.t., respectively. Among the 10122 and 21428 on-time events taken in 2008 and 2009, respectively 1698 and 3693 events were identified as candidate neutrino interactions in the target bricks.

\section{Event selection and analysis}
\label{selection}
The appearance of the $\tau$ lepton is identified in OPERA by the detection of its characteristic decay topologies, either in one prong (electron, muon or hadron) or in three prongs. 

A software algorithm  \cite{opcarac} selects  CC and neutral-current (NC) events inside the target. When a trigger in the electronic detectors is compatible with an interaction inside a brick, a software reconstruction program processes the electronic detector data to select the brick with the highest probability to contain the neutrino interaction vertex. The contamination by interactions occurring outside the target is evaluated by Monte Carlo simulations to be 4.8\%. Bricks  designated by the finding algorithm are removed from the target wall and exposed to X-rays for CS-to-brick alignment and the CS doublet is detached from the brick and developed underground. The corresponding CS films are then analysed  to validate the designated brick. The information of the CS is later used for a precise prediction of the position of the tracks in the most downstream films of the brick, helping the tracks scan-back in the subsequent vertex finding procedure. Validation of the selected brick is achieved if at least one track compatible with hits reconstructed in the electronic detectors is detected in the CS films. The efficiency of the brick finding procedure from the analysis of the 2008 run is found to be 77\% after the most probable or, when required, the two most probable bricks have been analysed. This efficiency reaches 83\% in a subsample where up to 4 bricks per event were processed.

After a brick has been validated, it is brought to the surface to be exposed to high-energy cosmic-rays for a precise film-to-film alignment. The brick emulsion films are then developed and dispatched to the various scanning laboratories. All tracks measured in the CS are sought in the most downstream films of the brick and followed back until they are not found in three consecutive films. The stopping point is considered as the signature either for a primary or a secondary vertex. The vertex is then confirmed by scanning a volume with a transverse size of 1 cm$^2$ in 15 films in total, 5 upstream and 10 downstream of the stopping point. The present overall location efficiency averaged over NC and CC events, from the electronic detector predictions down to the vertex confirmation, amounts to about 60\%.

A further phase of analysis is applied to located vertices, called decay search procedure, to detect possible decay or interaction topologies on tracks attached to the primary vertex. It is also searched for extra tracks from neutral decays, interactions and $\gamma$-ray conversions. When secondary vertices are found in the event, a kinematical analysis is performed, using particle angles and momenta measured in the emulsion films.

For charged particles up to about 6 GeV/c, momenta can be determined using the angular deviations produced by Multiple Coulomb Scattering (MCS) of tracks in the lead plates  \cite{MCS}. This method gives a momentum resolution better than 22\% for charged particles with momenta lower than 6 GeV/c, passing through an entire brick at normal incidence. For higher momentum particles, the measurement is based on the position deviations. The resolution is better than 33\% on 1/p up to 12 GeV/c for particles passing through an entire brick at normal incidence.  Momenta of muons reaching the spectrometer are measured with a resolution of 20\% up to 30 GeV/c and the sign of their charge is determined \cite{hpt}. 

The $\gamma$-ray search is performed in the whole scanned volume by checking all tracks having an impact parameter (IP) with respect to the primary or secondary vertices lower than 800 $\mu$m. The angular acceptance is   $\pm$500 mrad. The $\gamma$-ray energy is estimated by a Neural Network algorithm that uses the combination of the number of segments and the shape of the electromagnetic shower and also the multiple Coulomb scattering of the leading tracks. The same method is applicable to electrons from  $\nu_e$ CC interactions. The $\gamma$-ray attachment to a vertex is based on the angular resolution of the first measured segment of the $e^+e^-$  pair. The precision in the $\gamma$-ray attachment to a vertex depends on the conversion point inside the 1 mm thick lead plate. The typical resolution for $\gamma$-ray angle determination is 5 (3) mrad at 1 (3) GeV.

The detection of decay topologies is triggered by the observation of a track with a large impact parameter with respect to the primary vertex. The IP of primary tracks is smaller than 12 $\mu$m after excluding tracks produced by very soft particles.

The results presented in this paper come from the decay search analysis of a sample of 1088 events of which 901 are classified as CC interactions. This sample, taking into account the event location efficiencies, corresponds to 1813 interactions occurring in the target, namely about 35\% of the 2008 and 2009 data sample, or, rescaled to the beam integrated intensity, to 1.89$\times$10$^{19}$ p.o.t.. 

%\begin{figure}
%\begin{center}
%  % Requires \usepackage{graphicx}
% \includegraphics[width=14cm]{./ip.eps}
%   \caption{The histogram in red shows the MC simulation of the IP distribution of the tracks attached to the primary vertex and the dots represent real data. The histogram in blue shows the MC simulation of the IP distribution of the $\tau$ daughter with respect to that vertex.} \label{IP}
%\end{center}
%\end{figure}

Charmed particles have similar lifetimes as $\tau$ leptons and, if charged, share the same decay topologies. Being produced in $\nu_\mu$ CC interactions, the location of their primary vertex is easier than that of muon-less $\nu_\tau$ CC interactions where the $\tau$ lepton decays into hadrons or into an electron. Still, the primary vertex being located, the finding efficiency of the decay vertices is similar for both types of particles, although the selection criteria are more stringent in the case of the $\tau$ analysis that is aiming at minimal background. In the sample of  $\nu_\mu$ CC interactions that have been searched for, a total of 20 charm decays have been observed that survived selection cut-offs, in agreement with the number expected from a MC simulation, 16.0$\pm$2.9. This demonstrates that the efficiency of the search for short-lived decay topologies is understood. Among them, 3 have a one-prong topology where $0.8\pm0.2$ are expected. The background in the total charm events sample is about 2 events.

\section{Candidate event topology and track kinematics}
\label{topology}
The decay search procedure applied to the 2008-2009 data sample analysed so far yielded one candidate event with measured characteristics fulfilling the selection criteria a priori defined for the $\nu_\tau$ interaction search. The cuts used in this selection are those defined in the experiment proposal \cite{proposal1}  and its addendum  \cite{proposal2}. 

For decays of the $\tau$ to a single charged hadron, it is required at the secondary vertex, when not in the primary vertex lead plate, that:
\begin{itemize}
\item the kink angle $\theta_{kink}$ is larger than 20 mrad;
\item the secondary vertex is within the two lead plates downstream of the primary vertex;
\item the momentum of the charged secondary particles is larger than 2 GeV/c;
\item the total transverse momentum ($P_T$) of the decay products is larger than 0.6 GeV/c if there are no photons emitted at the decay vertex, and 0.3 GeV/c otherwise.
\end{itemize}

At the primary vertex, the selection criteria are:
\begin{itemize}
\item there are no tracks compatible with that of a muon or an electron;
\item the missing transverse momentum ($P_T^{miss}$) is smaller than 1 GeV/c;
\item the angle $\Phi$ in the transverse plane between the $\tau$ candidate track and the hadronic shower direction is larger than $\pi$/2.
\end{itemize}

The muon-less neutrino interaction reported in detail here occurred in a brick situated in wall 11 of the first SM and well inside the target: 3 bricks from the top and 24 bricks from the left. This allows a comprehensive study of the event: tracks can be followed over large distances and secondary vertices including electromagnetic showers searched for in a large volume. The event has been independently measured with the European and Japanese scanning systems.  Measurements are consistent and their averages are considered in the following.

\begin{figure}
\begin{center}
  \includegraphics[width=6.5cm]{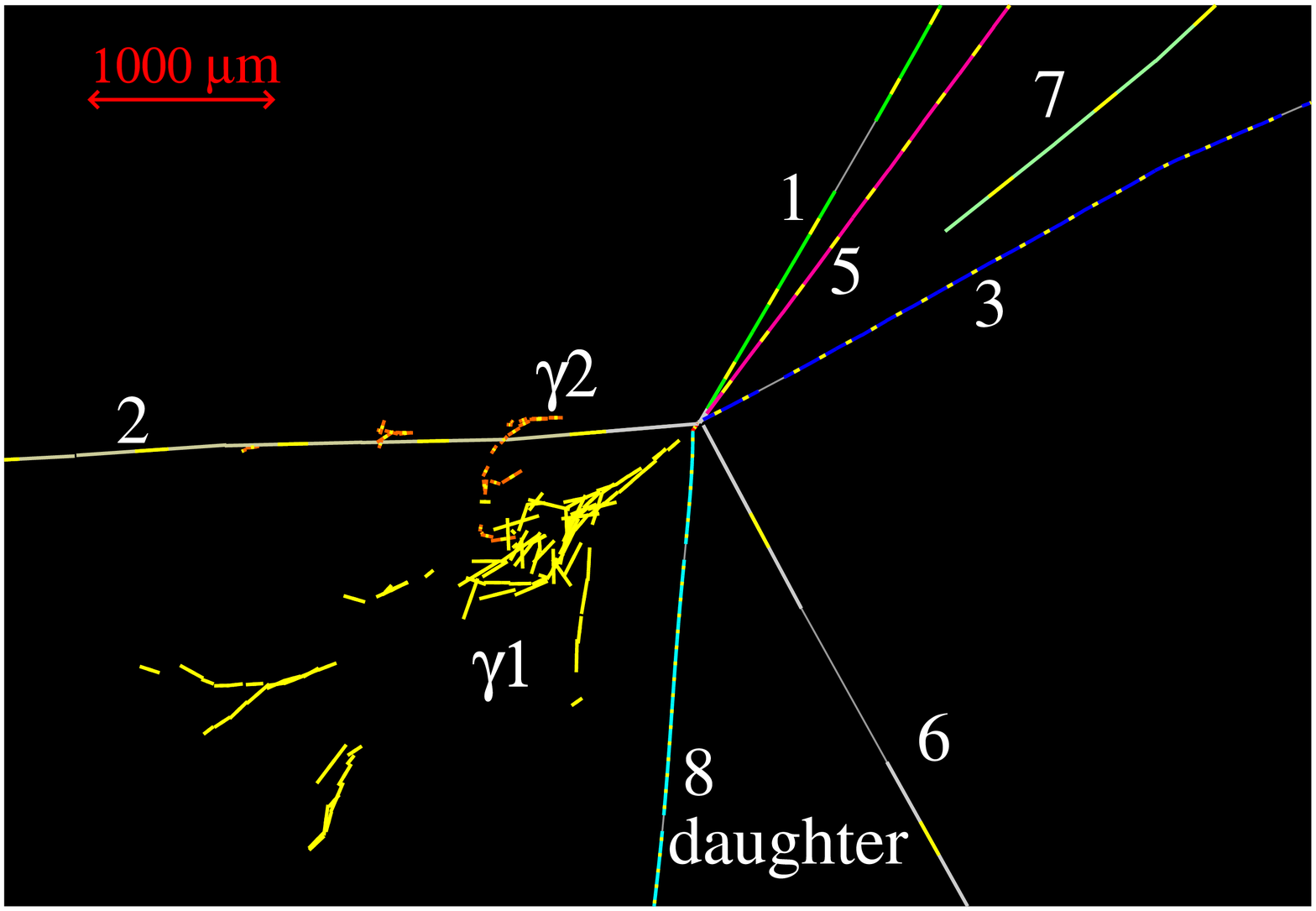}
    \includegraphics[width=6.5cm]{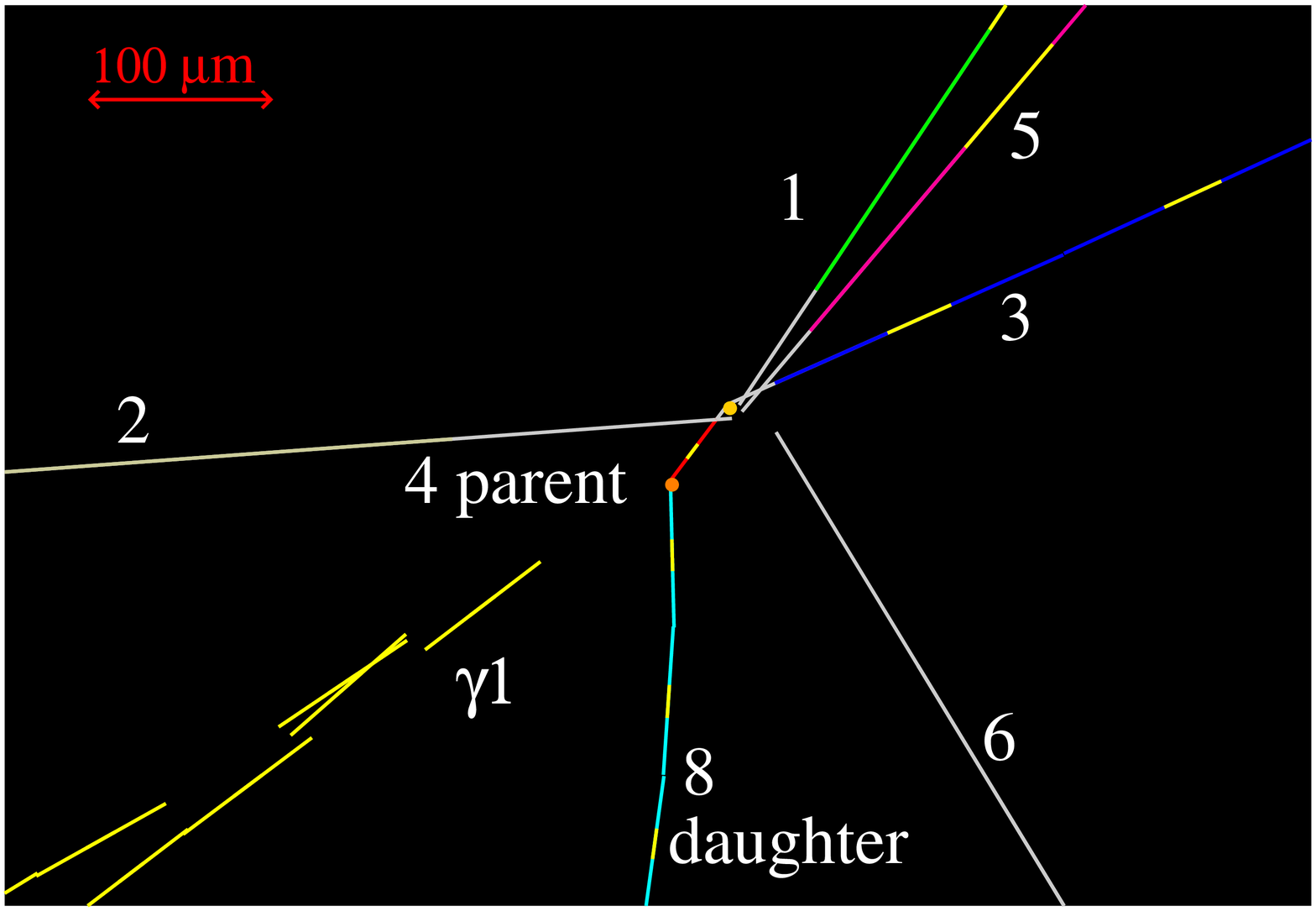}\\
      \includegraphics[width=13cm]{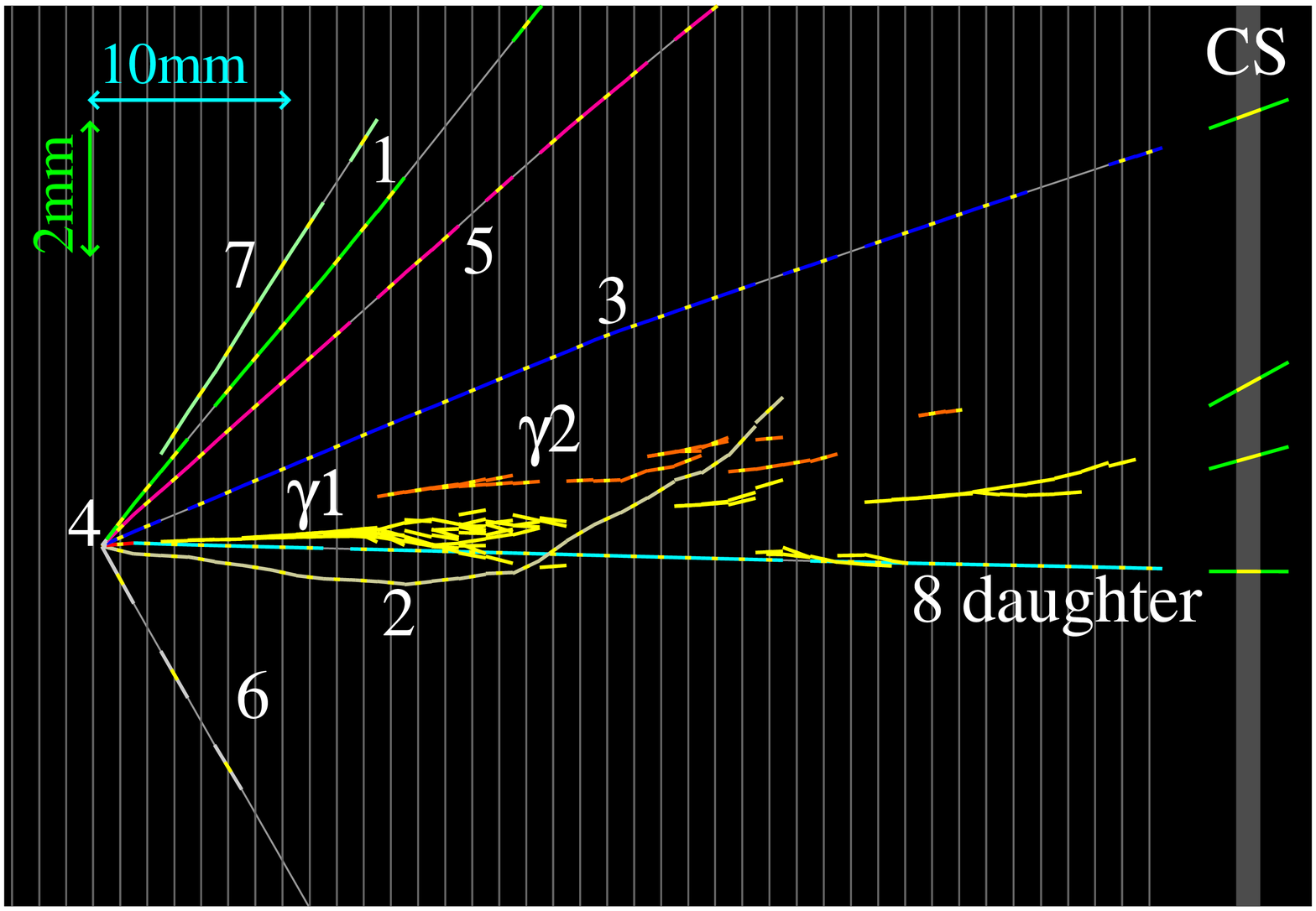}
   \caption{Display of the $\tau^-$ candidate event. Top left: view transverse to the neutrino direction. Top right: same view zoomed on the vertices. Bottom: longitudinal view.} \label{event}
\end{center}
\end{figure}

The primary neutrino interaction consists of 7 tracks of which one exhibits a visible kink. Two electromagnetic showers caused by $\gamma$-rays have been located that are associated with the event. Fig.  \ref{event} shows a display of the event.

\begin{figure}
\begin{center}
  \includegraphics[width=6.5cm]{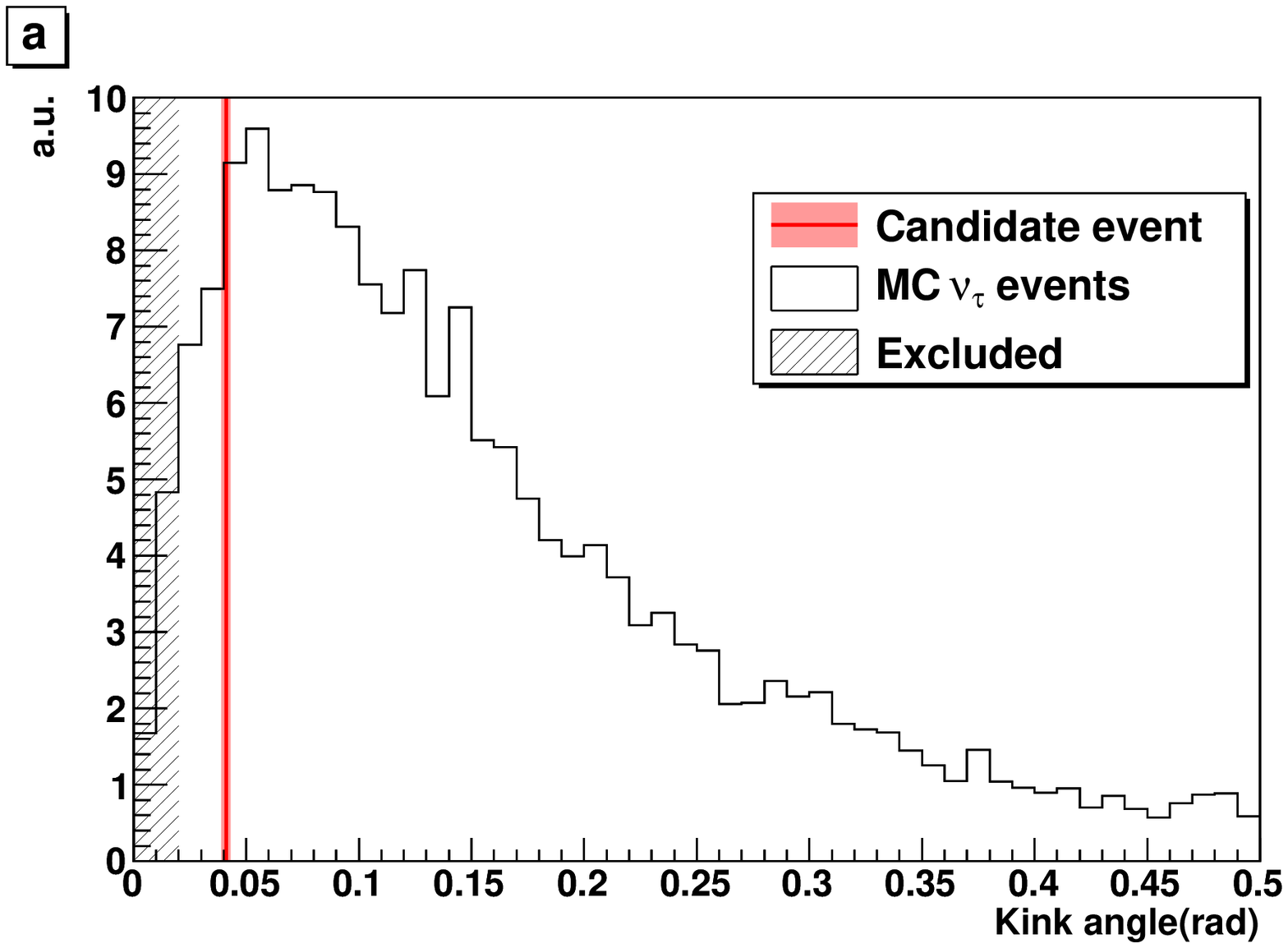}
   \includegraphics[width=6.5cm]{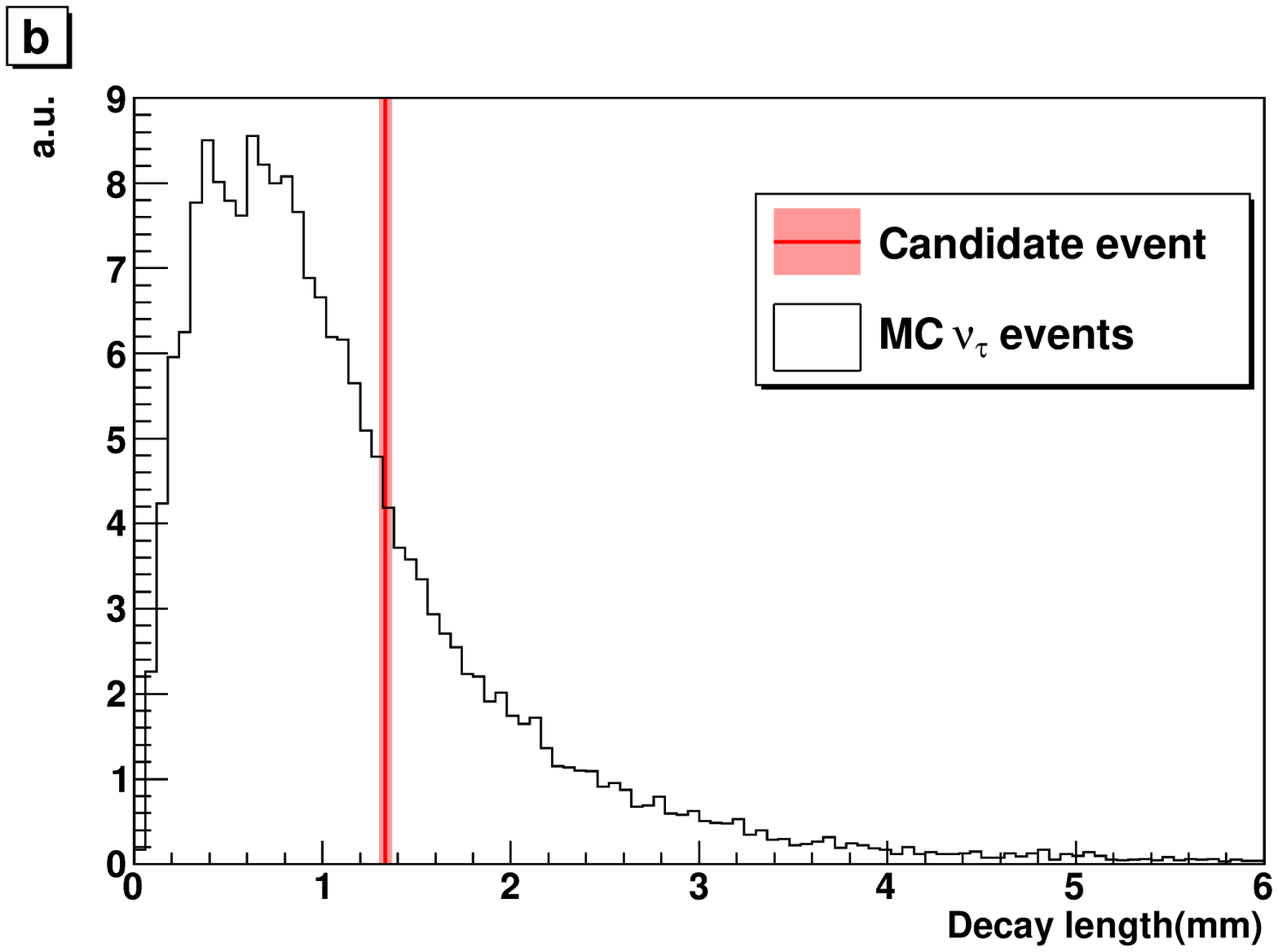}
   \includegraphics[width=6.5cm]{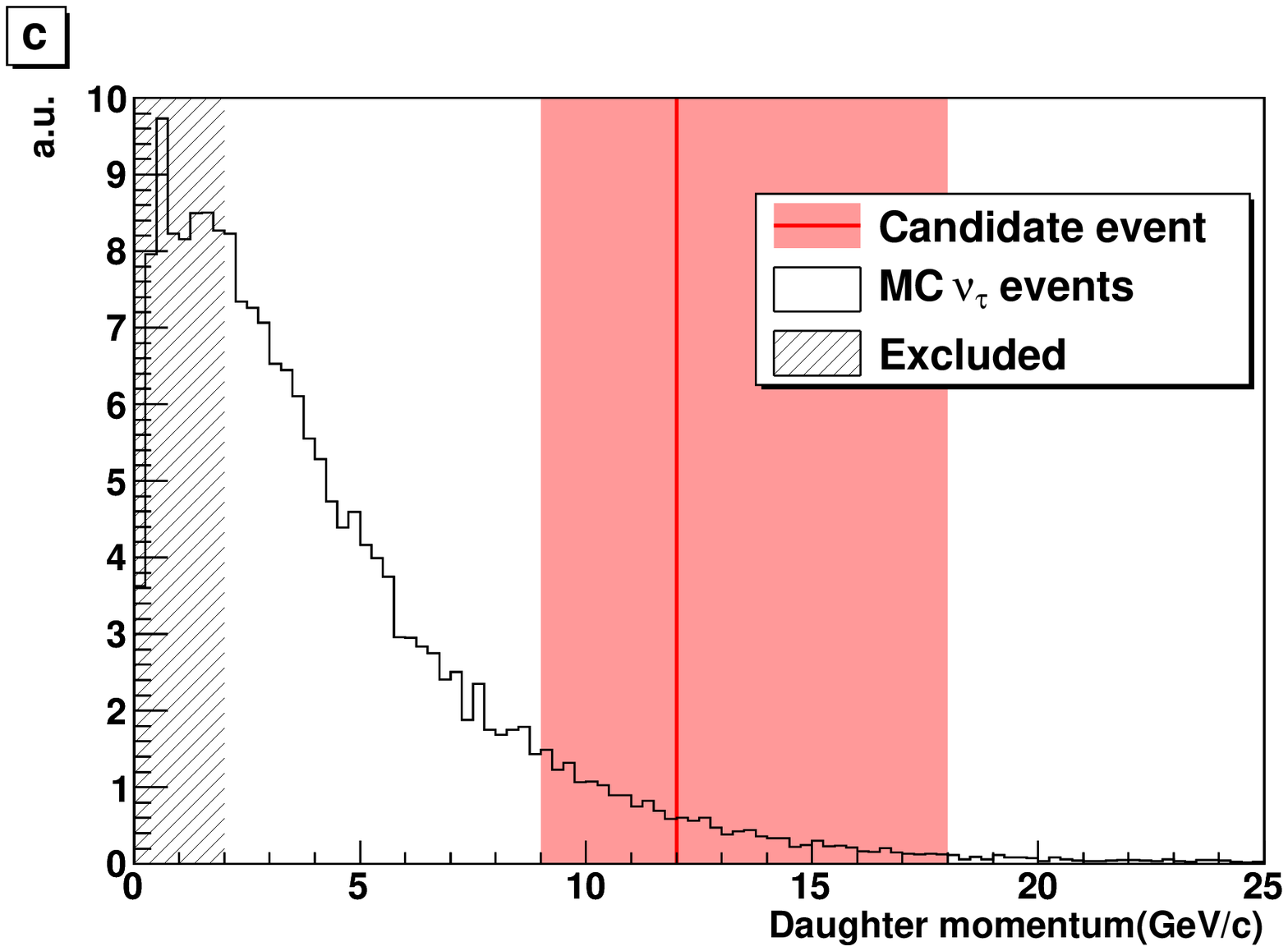}
    \includegraphics[width=6.5cm]{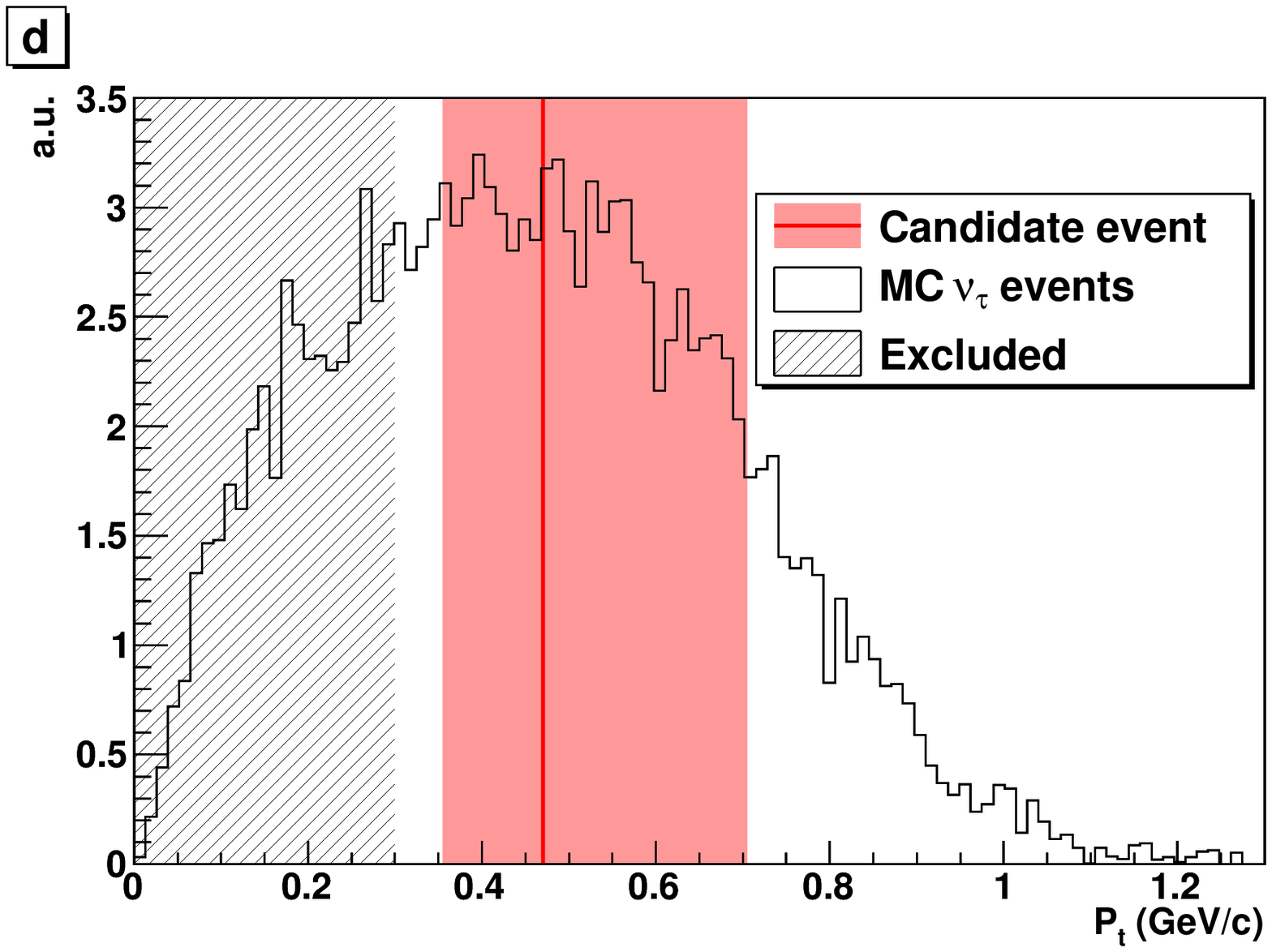}
   \caption{MC distribution of: a) the kink angle for $\tau$ decays. b) the path length of the $\tau$. c) the momentum of the decay daughter. d) the total transverse momentum $P_T$ of the detected daughter particles of $\tau$ decays with respect to the parent track. The red band shows the 68.3\% domain of values allowed for the candidate event and the dark red line the most probable value. The dark shaded area represents the excluded region corresponding to the a priori tau selection cuts.} \label{DK}
\end{center}
\end{figure}

\begin{itemize}
\item Track 1 exits from the primary interaction brick and is not found in the brick immediately downstream. It is left by a particle of momentum ($0.78^{+0.13}_{-0.10}$) GeV/c most likely interacting in the target tracker between both walls;
\item Track 2 is left by a heavily ionizing particle. From its residual range ($32.0\pm0.5$) g cm$^{-2}$ and the value of $p\beta$ = ($0.32^{+0.31}_{-0.11}$) GeV/c  measured on the upstream half of the track, the particle is identified as a proton, the kaon hypothesis being rejected with a C.L. of 97\%. The proton momentum resulting from the residual range corresponds to ($0.60\pm0.05$)  GeV/c;
\item Track 3 is left by a particle which generates a two-prong interaction 4 bricks downstream of the primary vertex. Its momentum is equal to ($1.97^{+0.33}_{-0.25}$) GeV/c;
\item Track 5 has been followed in wall 12 and disappears in wall 13 after a total distance shorter than 174  g cm$^{-2}$. The particle has a momentum of ($1.30^{+0.22}_{-0.16}$) GeV/c;
\item Track 6 stops in the primary brick. The range, ionisation and multiple scattering are that of a pion of very low momentum ($0.36^{+0.18}_{-0.09}$) GeV/c;
\item Track 7 is not directly attached to the primary vertex and points to it with an IP = ($43^{+45}_{-43}$) $\mu$m. Its starting point is separated from this vertex by 2 lead plates. Its origin is likely to be a prompt neutral particle. In the analysis, the momentum  ($0.49^{+0.29}_{-0.13}$) GeV/c has been added to the total momentum at the primary vertex;
\item Track 4 exhibits a kink topology with an angle of (41$\pm$2) mrad after a path length of (1335$\pm$35) $\mu$m corresponding to 15 $c\tau$ would it be that of a $\tau$ lepton. The expected $\gamma$ factor from the kink angle is approximately 25. It is the parent track of a secondary interaction or decay. Both the kink angle and the path length satisfy the selection criteria (Figs \ref{DK}a and \ref{DK}b);
\item  Track 8, the kink daughter track, is left by a particle  of a high momentum of ($12^{+6}_{-3}$) GeV/c well above the 2 GeV/c selection cut-off and which generates a 2-prong interaction 7 walls downstream its emission vertex (Fig. \ref{DK}c). Its IP with respect to the primary vertex is ($55\pm4$) $\mu$m. All the tracks directly attached to the primary vertex match the vertex point within 7  $\mu$m. 
\end{itemize}

None of the tracks of the charged particles emitted at either vertex is compatible with being that of an electron. Based on momentum-range consistency checks, the probability that tracks 1, 5 and 6 are left by muons is estimated to less than 0.001. The remaining tracks are identified as left by hadrons through their interaction. The probability of a soft muon with a slope larger than 1, which is undetectable in the brick and also in the TT, is 1.5\%.

$\gamma$-ray 1 has an energy of ($5.6\pm1.0(stat.)\pm 1.7(syst.)$) GeV. The distance between its conversion point and the secondary vertex is 2.2 mm and the shower is compatible with pointing to it with a probability of 32\%, the impact parameter being ($7.5\pm4.3$) $\mu$m. Its probability to be attached to the primary vertex is less than 10$^{-3}$, the IP being ($45.0 \pm 7.7$) $\mu$m.

$\gamma$-ray 2 has an energy of ($1.2\pm0.4(stat.)\pm0.4(syst.)$)  GeV. It is compatible with pointing to either vertex, with a significantly larger probability of 82\% at the secondary vertex, the IP being ($22^{+25}_{-22}$) $\mu$m, compared to 10\% and (85$\pm$38) $\mu$m at the primary vertex. Its distance to both vertices is about 13 mm.

 \section{Kinematical analysis of the candidate event}
 \label{kinematics}

In the most probable hypothesis $\gamma$-ray 2 is emitted at the secondary vertex. The total transverse momentum $P_T$ of the daughter particles with respect to the parent track is ($0.47^{+0.24}_{-0.12}$) GeV/c, above the lower selection cut-off at 0.3 GeV/c. In the low probability hypothesis where $\gamma$-ray 2 is attached to the primary vertex, the systematic shift in $P_T$ does not exceed 50 MeV/c. This small effect is included in Fig. \ref{DK}d.

\begin{figure}
\begin{center}
  \includegraphics[width=6.5cm]{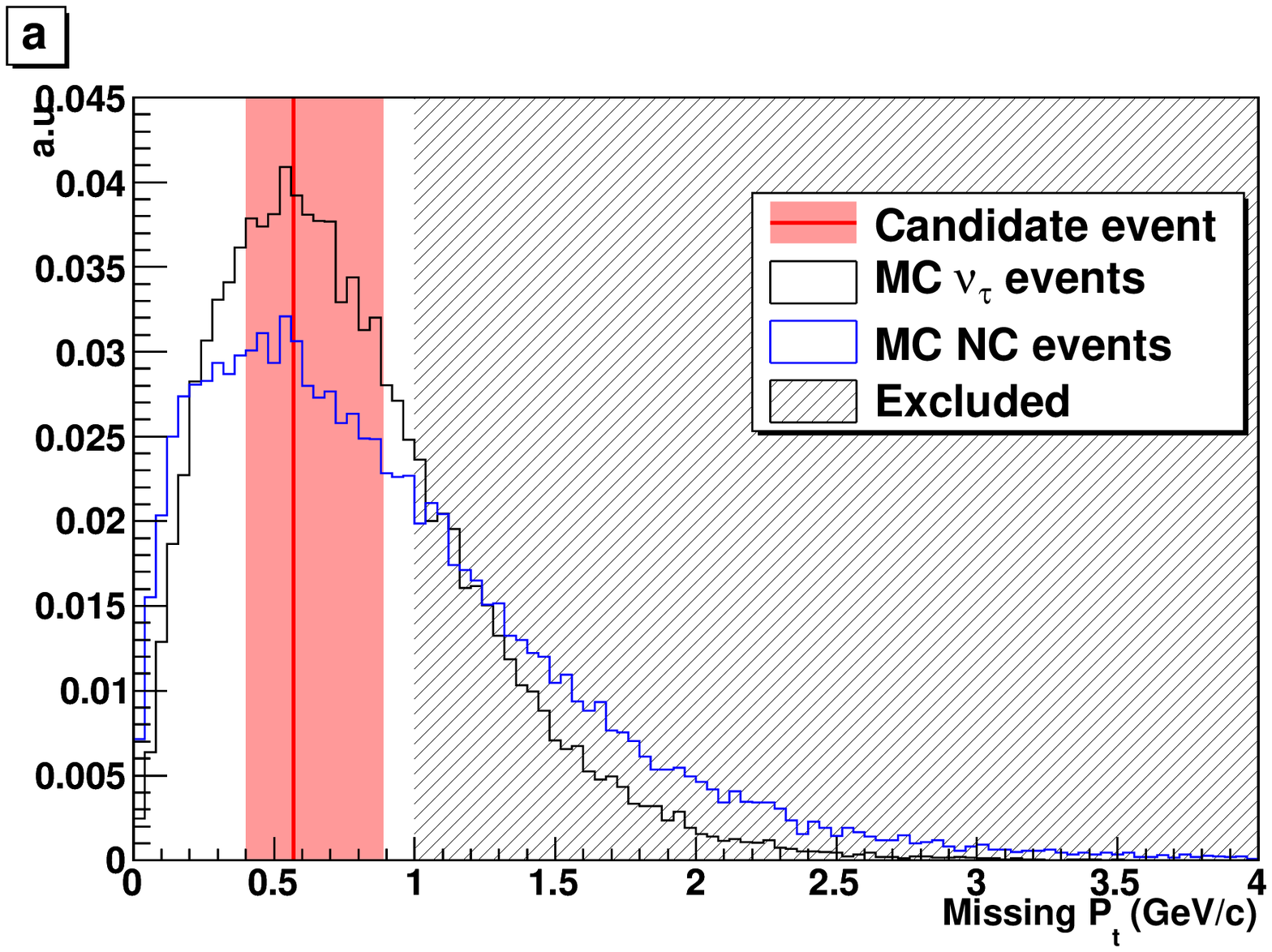}
    \includegraphics[width=6.5cm]{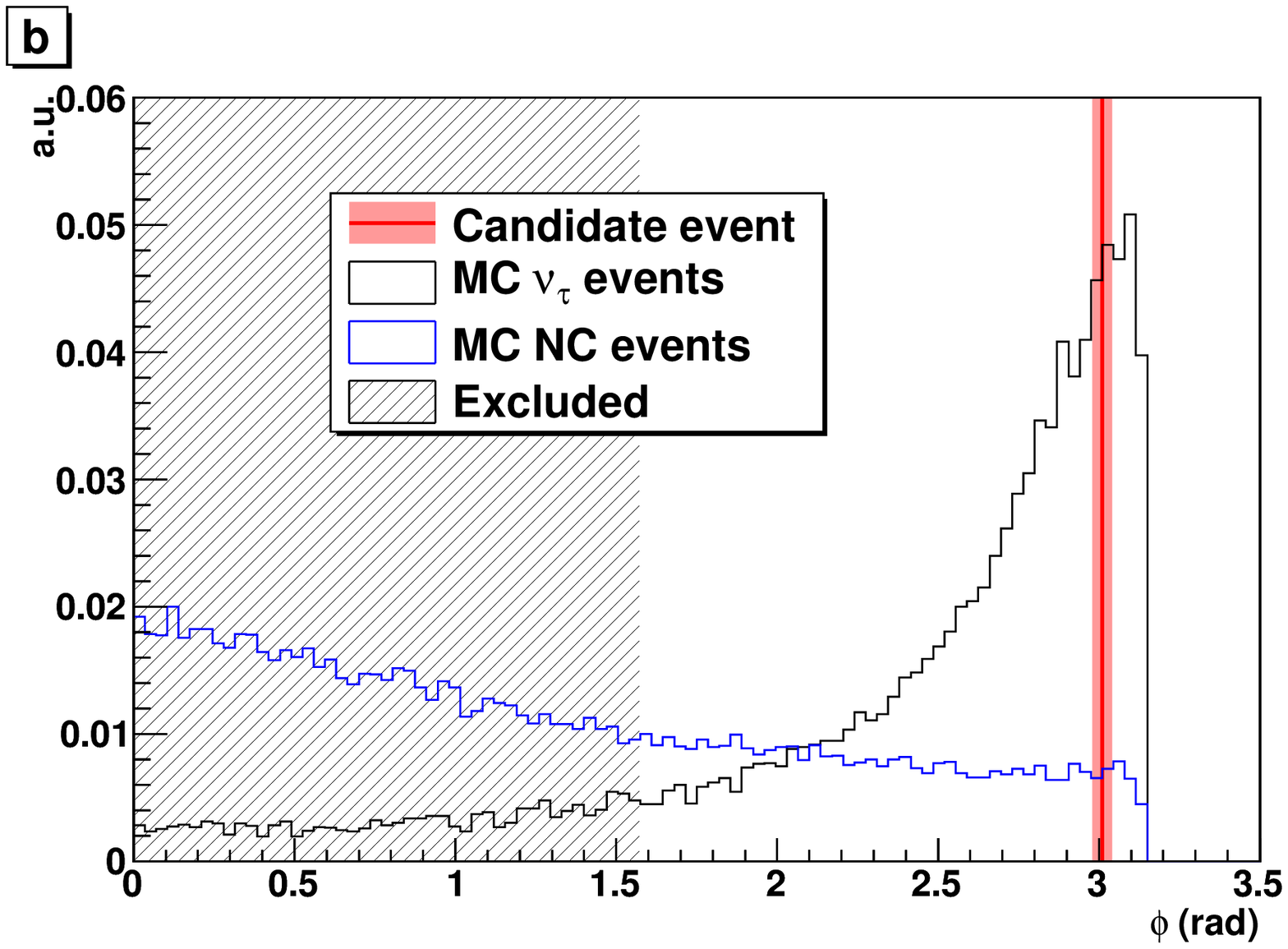}
   \caption{Monte-Carlo distribution of a) the missing $P_T$ at the primary vertex and b) the angle $\Phi$ between the parent track and the rest of the hadronic shower in the transverse plane at the primary vertex for $\nu_\tau$ CC interactions in black and for $\nu_\mu$ NC background in blue.  The red band shows the 68.3\% domain of values allowed for the candidate event and the dark red line the most probable value. The dark shaded area represents the excluded region corresponding to the a priori tau selection cuts.} \label{DK2}
\end{center}
\end{figure}

The missing transverse momentum $P_T^{miss}$ at the primary vertex is ($0.57^{+0.32}_{-0.17}$) GeV/c. This is lower than the upper selection cut-off at 1 GeV/c (Fig. \ref{DK2}a). The angle $\Phi$ between the parent track and the rest of the hadronic shower in the transverse plane is equal to ($3.01\pm0.03$) rad, largely above the lower selection cut-off fixed at $\pi$/2 (Fig. \ref{DK2}b). The total hadronic momentum at the primary vertex, not including that of the parent assumed to be a $\tau$ lepton, is about 5.5 GeV/c. The sum of the modulus of the momenta at the primary vertex is ($24.3^{+6.1}_{-3.2}$) GeV/c. This compares well with the total energy measured in the TT by calorimetry, (44$\pm$12) GeV, owing for the fraction of the momenta taken by the neutrals invisible in the bricks.

The invariant mass of $\gamma$-rays 1 and 2 is ($120\pm20(stat.)\pm35(syst.)$) MeV/c$^2$, supporting the hypothesis that they originate from a $\pi^0$ decay. Similarly the invariant mass of the charged decay product assumed to be a $\pi^-$ and of the two $\gamma$-rays amounts to ($640^{+125}_{-80}(stat.)^{+100}_{-90}(syst.)$) MeV/c$^2$, which is compatible with the $\rho(770)$ mass. The branching ratio of the decay mode $\tau\rightarrow\rho^-\nu_\tau$ is about 25\%.

\section{Background estimation and statistical significance}
\label{background}

The secondary vertex is compatible with the decay of a $\tau^-$ into $h^-(n\pi^0)\nu_\tau$. The two main sources of background to this channel where a similar final state may be produced are:
\begin{itemize}
\item the decays to a single charged hadron of charged charmed particles produced in $\nu_\mu$ CC interactions where the primary muon is not identified as well as the $c-\bar{c}$ pair production in $\nu_\mu$ NC  interactions where one charm particle is not identified and the other decays to a one-prong hadron channel;
\item the one-prong inelastic interactions of primary hadrons produced in $\nu_\mu$ CC interactions where the primary muon is not identified or in $\nu_\mu$ NC  interactions and in which no nuclear fragment can be associated with the secondary interaction.
\end{itemize}

The charm background produced in $\nu_\mu$ interactions in the analysed sample amounts to 0.007$\pm$0.004 (syst.) events \cite{proposal1,proposal2}. That produced in $\nu_e$ interactions is less than 10$^{-3}$ events. Current estimates of charm background are conservative since they are based on the original scanning strategy of the proposal. They do not include the additional reduction obtained by following all tracks up to their end points as was done for this candidate event. An improved evaluation of the charm background that exploits such additional information is in progress.

A search was performed for hadronic activity seeking for highly ionising tracks associated with nuclear fragments and pointing to the secondary vertex.  No such track is observed while the number of $\alpha$-ray tracks in the vicinity of the event is found to be compatible with that expected from the lead natural radioactivity \cite{facilities}. The dominant background from hadron re-interactions has been evaluated with a FLUKA based Monte Carlo code \cite{FLUKA}. About 160 millions of events were simulated where 0.5-15 GeV $\pi^+$, $\pi^-$, $K^+$, $K^-$ and $p$ are impinging on 1 mm of lead. The probability for a background interaction to occur over 2 mm of lead, the maximal decay length considered, and to satisfy the selection criteria for the reconstruction of the kink decay topology and its kinematics is $(1.9\pm 0.1)\times10^{-4}$ per NC event. This probability decreases to $(3.8\pm 0.2)\times10^{-5}$  per NC event when taking into account the cuts on the event global kinematics. This leads to a total of  $0.011\pm0.006 (syst.)$ background events when misclassified CC events are included. We assume a 50\% error on the result of the FLUKA simulation in order to take into account possible systematic effects in the hadronic interactions modeling.

The total background in the decay channel to a single charged hadron is therefore $0.018\pm0.007 (syst.)$. This new estimation based on an upgraded and more reliable simulation tool is 2.2 times larger than the value of the experiment proposal  \cite{proposal1, proposal2}.  
                                                                                                                          
Three experimental cross-checks of the one-prong hadron background calculation have been performed, though so-far with small statistics. The tracks of hadrons from a sub-sample of NC and CC interactions have been followed far from the primary vertex over a total length of 8.6 m and no background-like interaction has been found in the signal region. This corresponds to a probability of occurrence over 2 mm of lead smaller than  $1.54\times10^{-3}$ at 90\% C.L. Outside the signal region, a total of 7 single-prong interactions have been located with a kink angle $>$ 20 mrad and a daughter transverse momentum $P_T>0.2$  GeV/c while 8 are expected from the simulation. Within the low statistics, both $P_T$ distributions agree well. Finally, a brick has been exposed to a 4 GeV/c $\pi$- beam and 119 interactions have been analysed so far. The forward and backward track multiplicities, the kink angle, the secondary particles momentum and their $P_T$ have been checked against FLUKA. They agree within statistical uncertainties both in normalisation to $\sim$10\% and shape. $(7.4\pm1.1)$\% are one-prong interactions with a kink angle larger than 20 mrad where $(7.6\pm0.3)$\%  are predicted by the simulation.

Finally, the expected number of events due to prompt $\nu_\tau$ background from $D_s$ decays is 10$^{-4}$ (10$^{-7}$ per CC event).

The probability that the expected $0.018\pm0.007$(syst.) background events to the $h^-(n\pi^0)\nu_\tau$ decay channel of the $\tau$ fluctuate to one event is 1.8\%  (2.36 $\sigma$). 
As the search for $\tau$- decays is extended to all four channels,  $\mu^-\nu_\tau\bar{\nu}_\mu$, $e^-\nu_\tau\bar{\nu}_e$, $h^-(n\pi^0)\nu_\tau$ and  $\pi^-\pi^-\pi^+\nu_\tau$, the total background then becomes $0.045\pm0.023 (syst.)$. The probability that this expected background to all searched decay channels of the $\tau$ fluctuate to one event is 4.5\%  (2.01 $\sigma$). 

At $\Delta m^2_{23}=2.5\times10^{-3}$ eV$^2$ and full mixing, the expected number of observed $\tau$ events with the present analysed statistics is $0.54\pm0.13 (syst.)$ of which $0.16\pm0.04 (syst.)$ in the one-prong hadron topology, in agreement with the observation of one event.

\section{Conclusions}
\label{conclu}
We reported the observation of a first candidate $\nu_\tau$ CC interaction in the OPERA detector at LNGS. It was identified in a sample of events corresponding to $1.89\times10^{19}$ p.o.t. in the CERN CNGS $\nu_\mu$ beam. The assumed $\tau^-$ lepton decays into $ h^- (n\pi^0)\nu_\tau$ and passes the selection criteria \cite{proposal1, proposal2}. 

The observation of one possible tau candidate in the decay channel $h^-(\pi^0)\nu_\tau$ has a significance of 2.36 $\sigma$ of not being a background fluctuation. If one considers all decay modes included in the search the significance of the observation becomes 2.01 $\sigma$. The expected number of $\nu_\tau$ events detected in the analysed sample is $0.54\pm0.13 (syst.)$.

We therefore claim the observation of a first candidate $\nu_\tau$ CC interaction. Its significance, based on our best conservative knowledge of the background, exceeds two $\sigma$. This does not allow yet claiming the observation of $\nu_\mu\rightarrow\nu_\tau$ oscillation. Given its sensitivity, the OPERA experiment will require the detection of a few more candidate events in order to firmly establish neutrino oscillations in direct appearance mode through the identification of the final charged lepton. 

%In parallel we are in the process of further improving the knowledge of the different efficiencies involved in the analysis procedure and increasing the experiment sensitivity using data from the accumulating statistics of the reconstructed neutrino events. This will provide a better estimation of the backgrounds due to hadronic re-interactions and charm decays and eventually allow further optimisation of the selection cuts.

\section{Acknowledgements}
We thank CERN for the commissioning of the CNGS facility and for its successful operation and INFN for the continuous support given to the experiment during the construction, installation and commissioning phases through its LNGS laboratory. We warmly acknowledge funding from our national agencies: Fonds de la Recherche Scientifique - FNRS and Institut Interuniversitaire des Sciences Nucl{\'e}aires for Belgium, MoSES for Croatia, CNRS and IN2P3 for France, BMBF for Germany, INFN for Italy, JSPS (Japan Society for the Promotion of Science), MEXT (Ministry of Education, Culture, Sports, Science and Technology ), QFPU (Global COE program of Nagoya University, "Quest for Fundamental Principles in the Universe" supported by JSPS and MEXT) and Promotion and Mutual Aid Corporation for Private Schools of Japan, SNF and ETH Zurich for Switzerland, the Russian Federal Property Fund, the grant 09-02-00300\_a, 08-02-91005-CERN, Programs of the Presidium of the Russian Academy of Sciences ''Neutrino Physics'' and ''Experimental and theoretical researches of fundamental interactions connected with work on the accelerator of CERN'', Programs of support of leading schools of thought  (grant 3517.2010.2), Federal Agency on a science and innovations state contract 02.740.11.5092 for Russia., the Korea Research Foundation Grant (KRF-2008-313-C00201) for Korea. We are also indebted to INFN for providing fellowships and grants to non Italian researchers. We thank the IN2P3 Computing Centre (CC-IN2P3) for providing computing resources for the analysis and hosting the central database for the OPERA experiment. We are indebted to our technical collaborators for the excellent quality of their work over many years of design, prototyping and construction of the detector and of its facilities. Finally, we thank our industrial partners.



%% References


%\bibliographystyle{elsarticle-num}

\begin{thebibliography}{00}


\bibitem{pontecorvo}
B. Pontecorvo, J.Exptl. Theoret. Phys. $\mathbf{33}$ (1957) 549) and Sov. Phys. JETP $\mathbf{6}$ (1957) 429;\\
B. Pontecorvo, J.Exptl. Theoret. Phys. $\mathbf{34}$ (1958) 247 and Sov. Phys. JETP $\mathbf{7}$ (1958) 172;\\
Z. Maki, M. Nakagawa and S. Sakata, Prog. Theor. Phys. $\mathbf{28}$ (1962) 870.\\

\bibitem{SK}
SUPER-KAMIOKANDE collaboration, Y. Fukuda et al., Phys. Rev. Lett. $\mathbf{81}$ (1998) 1562;\\
SUPER-KAMIOKANDE collaboration, K. Abe et al., Phys. Rev. Lett. 97 (2006) 171801;\\
SUPER-KAMIOKANDE collaboration, R. Wendell et al., arXiv:1002.3471.


\bibitem{review}
C. Amsler et al. (Particle Data Group), Phys. Lett. B 667 (2008) 1;\\
See 2009 updates at http://pdg.web.cern.ch/pdg/

\bibitem{K}
KAMIOKANDE-II collaboration, K.S. Hirata et al., Phys. Lett. B 205 (1988) 416.

\bibitem{macro}
MACRO collaboration,M. Ambrosio et al., Eur. Phys. J. C 36 (2004) 323.

\bibitem{soudan2}
SOUDAN-2 collaboration, W.W.M. Allison et al., Phys. Rev. D 72 (2005) 052005.

\bibitem{chooz}
CHOOZ collaboration, M. Apollonio et al., Eur. Phys. J. C 27 (2003) 331.

\bibitem{paloverde}
Palo Verde Collaboration, A. Piepke, Prog. Part. Nucl. Phys. 48 (2002) 113.

\bibitem{K2K}
K2K collaboration, M.H. Ahn et al., Phys. Rev. D 74 (2006) 072003.

\bibitem{MINOS}
MINOS collaboration, D.G. Michael et al., Phys. Rev. Lett. 97 (2006) 191801;\\
MINOS collaboration, P. Adamson et al., Phys. Rev. Lett. 101 (2008) 221804.

\bibitem{operaold}
A. Ereditato, K. Niwa and P. Strolin, The emulsion technique for short, medium and long baseline $\nu_\mu\rightarrow\nu_\tau$ oscillation experiments, 423, INFN-AE-97-06, DAPNU-97-07;\\
%A. Ereditato, K. Niwa and P. Strolin, Nucl. Phys. B (Conf. Suppl.) 66 (1998) 423;\\
OPERA collaboration, H. Shibuya et al., Letter of intent: the OPERA emulsion detector for a long-baseline neutrino-oscillation experiment, CERN-SPSC-97-24, LNGS-LOI-8-97.

\bibitem{proposal1}
OPERA collaboration, M. Guler et al., An appearance experiment to search for $\nu_\mu\rightarrow\nu_\tau$ oscillations in the CNGS beam: experimental proposal, CERN-SPSC-2000-028,  LNGS P25/2000

\bibitem{proposal2}
OPERA collaboration, M. Guler et al., Status Report on the OPERA experiment, CERN/SPSC 2001-025, LNGS-EXP 30/2001 add. 1/01

\bibitem{CNGS}
Ed. K. Elsener,  ÒThe CERN Neutrino beam to Gran Sasso (Conceptual Technical Design)Ó, CERN 98-02, INFN/AE-98/05;\\ 
R. Bailey et al., ÒThe CERN Neutrino beam to Gran Sasso (NGS)Ó (Addendum to report CERN 98-02, INFN/AE-98/05)Ó, CERN-SL/99-034(DI), INFN/AE-99/05.

\bibitem{operafirst}
OPERA collaboration, R. Acquafredda et al., New J. Phys. 8 (2006) 303;\\
OPERA collaboration, N. Agafonova et al.,JINST 4 (2009) P06020.

\bibitem{niu}
K. Niu, E. Mikumo and Y. Maeda, Prog. Theor. Phys. 46 (1971) 1644.

\bibitem{donut}
DONUT collaboration, K. Kodama et al., Phys. Lett. B 504 (2001) 218;\\
DONUT collaboration, K. Kodama et al., Nucl.Instrum.Meth. A 493 (2002) 45.

\bibitem{operadetector}
OPERA collaboration, R. Acquafredda et al., JINST 4 (2009) P04018.

\bibitem{ESS}
N. Armenise et al., Nucl. Instrum. Meth. A 551 (2005) 261;\\
M. De Serio et al., Nucl. Instrum. Meth. A 554 (2005) 247;\\
L. Arrabito et al.,  Nucl. Instrum. Meth. A 568 (2006) 578.

\bibitem{SUTS}
K. Morishima and T. Nakano, JINST (2010) 5 P04011.

\bibitem{facilities}
OPERA collaboration, A. Anokhina et al., JINST 3 (2008) P07002;\\
OPERA collaboration, A. Anokhina et al., JINST 3 (2008) P07005;\\
T. Adam et al., 	Nucl. Instrum. Meth. A 577 (2007) 523;\\
T. Nakamura et al., Nucl. Instrum. Meth. A 556 (2006) 80.

\bibitem{opcarac}
A. Bertolin and N. T. Tran, OpCarac: an algorithm for the classification of the neutrino interactions recorded by OPERA, OPERA note 100 (2009).

\bibitem{MCS}
M. De Serio at al., Nucl. Instrum. Meth. 512 (2003) 539;\\
M. Besnier, PhD thesis (2008) UniversitŽ de Savoie, LAPP-T-2008-02, http://hal.in2p3.fr/tel-00322932\_v1 .

\bibitem{hpt}
R. Zimmermann et al., Nucl. Instrum. Meth. A 555 (2005) 435;\\
R. Zimmermann et al., erratum, Nucl. Instrum. Meth. A 557 (2006) 690.

\bibitem{FLUKA}
FLUKA, http://www.fluka.org/fluka.php;\\
G. Battistoni, et al. , Proceedings of the Hadronic Shower Simulation Workshop 2006, Fermilab 6-8 September 2006, M. Albrow, R. Raja eds., AIP Conference Proceeding 896, 31-49, (2007).

\end{thebibliography}



\end{document}